\documentclass{article}

\usepackage{arxiv}

\usepackage[utf8]{inputenc} 
\usepackage[T1]{fontenc}    
\usepackage{hyperref}       
\usepackage{url}            
\usepackage{booktabs}       
\usepackage{amsfonts}       
\usepackage{nicefrac}       
\usepackage{microtype}      
\usepackage{lipsum}		
\usepackage{graphicx}
\usepackage{natbib}
\usepackage{doi}

\usepackage{amsmath}
\usepackage{nicefrac}
\usepackage{xcolor}
\usepackage{bm}
\renewcommand{\vec}[1]{\bm{#1}}
\newcommand{\ten}[1]{\bm{#1}}
\usepackage[ruled]{algorithm2e}

\title{A Lattice Boltzmann Method for nonlinear solid mechanics in the reference configuration}

\date{August 25, 2022}	

\author{ \href{https://orcid.org/0000-0003-2189-945X}{\includegraphics[scale=0.06]{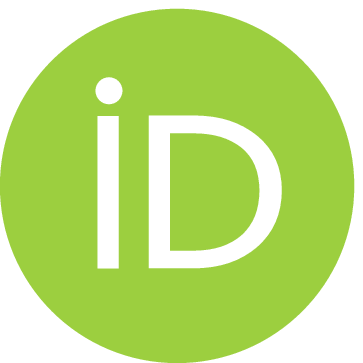}\hspace{1mm}Erik Faust}\thanks{Corresponding author.} \\
	Lehrstuhl für Technische Mechanik\\
	Technische Universität Kaiserslautern\\
	Postfach 3049, D-67653 Kaiserslautern \\
	\texttt{efaust@rhrk.uni-kl.de} \\
	\And
	\href{https://orcid.org/0000-0002-0839-6519}{\includegraphics[scale=0.06]{orcid.eps}\hspace{1mm}Alexander Schlüter} \\
	Lehrstuhl für Technische Mechanik\\
	Technische Universität Kaiserslautern\\
	Postfach 3049, D-67653 Kaiserslautern \\
	\texttt{aschluet@rhrk.uni-kl.de} \\
	\And
	\href{https://orcid.org/0000-0003-4819-2198}{\includegraphics[scale=0.06]{orcid.eps}\hspace{1mm}Henning Müller} \\
	Institut für Mechanik\\
	Technische Universität Darmstadt\\
	Franziska-Braun-Straße~7, D-64287, Darmstadt \\
	\texttt{henning.mueller@tu-darmstadt.de} \\
	\And
	{\hspace{1mm}Ralf Müller} \\
	Institut für Mechanik\\
	Technische Universität Darmstadt\\
	Franziska-Braun-Straße~7, D-64287, Darmstadt \\
	\texttt{ralf.mueller@mechanik.tu-darmstadt.de} \\
}

\renewcommand{\shorttitle}{An LBM for nonlinear solid mechanics}

\hypersetup{
pdftitle={An LBM for nonlinear solid mechanics},
pdfsubject={q-bio.NC, q-bio.QM},
pdfauthor={Erik Faust, Alexander Schlüter, Henning Müller, Ralf Müller},
pdfkeywords={Lattice Boltzmann Method, Nonlinear Solid Mechanics, Large Deformation, Reference Configuration, Transient Simulation},
}

\begin{document}

\maketitle

\begin{abstract}
With a sufficiently fine discretisation, the Lattice Boltzmann Method (LBM) mimics a second order Crank-Nicolson scheme for certain types of balance laws \cite{farag}.
This allows the explicit, highly parallelisable LBM to efficiently solve the fundamental equations of solid mechanics: the conservation of mass, the balance of linear momentum, and constitutive relations.

To date, all LBM algorithms for solid simulation -- see e.g. \cite{murthy}, \cite{escande}, \cite{schlueter2} -- have been limited to the small strain case.
Furthermore, the typical interpretation of the LBM in the current (Eulerian) configuration 
is not easily extensible to large strains, as large topological changes complicate the treatment of boundary conditions.

In this publication, we propose a large deformation Lattice Boltzmann Method for geometrically and constitutively nonlinear solid mechanics. 
To facilitate versatile boundary modelling, the algorithm is defined in the reference (Lagrangian) configuration. 
\end{abstract}


\section{Introduction}

The Lattice Boltzmann Method (LBM) is a staple solver for simulations in fluid dynamics. It is particularly attractive on account of its computational efficiency and ease of discretisation \cite[p.55]{krueger}. 
Recently, this efficiency and the promise of coupled fluid-structure simulations using the LBM has also driven a rising research interest into Lattice Boltzmann (LB) schemes for solid simulation \cite{murthy,escande}.

\cite{marconi,obrien,schlueter2}, for example, proposed LB algorithms of increasing generality based on wave equation formulations for the Lam\'e-Navier equation. \cite{marconi,schlueter1,mueller} used similar schemes for simulations in fracture mechanics. Finally, \cite{murthy} and \cite{escande} developed a solid-mechanical LBM based on balance laws in the small strain setting.
These LB algorithms allow for efficient simulations of transient phenomena in linear elastic continua. Additional developments in boundary rules -- see e.g. \cite{schlueter1,schlueter3,us} -- have also made it possible to model the behaviour of solid bodies with more complicated geometries.

Until now, all such LB algorithms for solid simulation have been limited to small deformations.
Furthermore, the typical interpretation of the LBM in the current (Eulerian) configuration is not easily extensible to large strains, as large topological changes complicate the treatment of boundary conditions.

An extension to large deformations might open up new fields of application for the LBM in solid mechanics: impact simulations and biomechanical tissue modelling, for example, may eventually become feasible. Additionally, nonlinear material models could allow for simulations of tires, seals, and other polymer components under transient load. A formulation in the reference configuration, meanwhile, could greatly simplify the modelling of boundaries in certain situations with large deformations.
Furthermore, it may also become feasible to use the LBM to model both the fluid and the solid behaviour in coupled fluid-structure simulations.

This work proposes a large displacement Lattice Boltzmann Method for geometrically and materially nonlinear solid mechanics in the reference configuration. The publication is organised as follows:
section \ref{s:intro} briefly motivates why the LBM can be used as a solver for balance laws in the reference configuration, based on a recent numerical investigation of the LBM by \cite{farag}.
Section~\ref{s:LBMrefcon} then outlines an LB scheme from the Lagrangian point of view, and section~\ref{s:mc} presents an appropriate moment chain for nonlinear solid mechanics. In section~\ref{s:FSMC_LBM}, an LB algorithm is proposed to solve this moment chain, and section~\ref{s:BC} covers the boundary rules used. 
Finally, results obtained from the LBM are validated against FEM reference solutions for two simple examples and one more challenging benchmark. 

\section{The LBM as a solver for moment chains}\label{s:intro}

As demonstrated recently \cite{farag}, the Lattice Boltzmann Method (LBM) mimics a second-order Crank-Nicolson scheme for balance laws of the form
\begin{equation}\label{eq:chain1}
    \frac{d}{dt} \Big ( {}^N \Pi_{[\alpha]} \Big ) + \frac{d}{dx_\beta} \Big ( {}^{N+1} \Pi_{[\alpha]\beta} \Big ) = {}^N S_{[\alpha]}\,,
\end{equation}
where $\ten{{}^N \Pi}$ represents some $N$-th order tensor field and $[\alpha]$ designates a set of indices $\alpha_1,...,\alpha_N$, while $\ten{{}^N S}$ is an $N$-th order source term. This is a profound realisation: for sufficiently small time steps, the explicit, highly parallelisable LBM acts as an implicit Runge-Kutta method in moment space.
As it uses purely numerical arguments, the derivation in \cite{farag} further generalises the LBM beyond the spatial particle transport context of the Boltzmann equation \cite[p.21]{krueger} in which it is usually viewed. 
Consequently, we can interpret the temporal and spatial derivatives in (\ref{eq:chain1}) as being defined in the reference (Lagrangian) configuration, and use the LBM to solve conservation laws of the type
\begin{equation}\label{eq:chainRK}
    d_t {}^N \Pi_{[\alpha]} + d_\beta {}^{N+1} \Pi_{[\alpha]\beta} = {}^N S_{[\alpha]}\,,
\end{equation}
where $d_t$ and $d_\alpha$ are the material time derivative and material gradient/divergence operators, respectively.
When unravelled, the first three `chain links' of this moment chain could also be written as
\begin{align}\label{eq:chain}
    & d_t {}^0 \Pi + d_\alpha {}^{1} \Pi_{\alpha} = {}^0 S\,,\nonumber\\
    & d_t {}^1 \Pi_{\alpha} + d_\beta {}^{2} \Pi_{\alpha\beta} = {}^1 S_{\alpha}\,,\nonumber\\
    & d_t {}^2 \Pi_{\alpha \beta} + d_\gamma {}^{3} \Pi_{\alpha\beta\gamma} = {}^2 S_{\alpha\beta}\,.
\end{align}
Equation (\ref{eq:chain}) describes balance laws with respect to material rather than spatial coordinates, which is more convenient for solid simulations, especially at large deformations. Solids do generally not assume the shape of vessels containing them. Thus, boundary conditions must be formulated with respect to parts of the material rather than regions in space. Furthermore, the material time derivatives often appearing in solid-mechanical balance laws yield convective terms in the Eulerian framework, but not in the Lagrangian setting \cite[p.64-69]{holzapfel}. This fact simplifies derivations considerably.

\section{The LBM (in the reference configuration)}\label{s:LBMrefcon}

This section briefly discusses the central steps of the LBM algorithm, which is formulated in the reference configuration here. The interested reader is referred to \cite{krueger} for an excellent textbook covering the LBM in detail.

We introduce a lattice consisting of lattice sites $\vec{X}\in L \subset \Omega_0$ in the reference configuration. The sites are spaced regularly at distances of $\Delta X$ in each coordinate direction. For the sake of brevity, only two-dimensional D2Q9 lattices are considered in this work. In such a lattice, each lattice site is connected to its $8$ nearest (direct and diagonal) neighbours via $8$ lattice links.
Furthermore, a set of $9$ lattice velocities\footnote{Note that this includes one zero velocity $\vec{C_i}=\vec{0}$.} $\vec{C_i}$ are defined to cover the distance between one lattice site $\vec{X}$ and its neighbours $\vec{X}+\vec{C_i}\Delta t$ in one time step $\Delta t$ \cite[p.86-88]{krueger}.

The LBM performs explicit operations on distribution functions $\bar{f}_i$ (one per lattice velocity $\vec{C_i}$) on this lattice: in each iteration, these are first relaxed towards the value of the equilibrium distribution function $f_i^{eq}$ with relaxation time $\bar{\tau}$,
\begin{equation}\label{eq:col}
    \bar{f}_i^{col} = \bar{f}_i - \frac{\Delta t}{\bar{\tau}}( \bar{f}_i - f_i^{eq} ) + \Delta t \Big (1 - \frac{\Delta t}{2 \bar{\tau}} \Big ) \psi_i\,,
\end{equation}
and a source term contribution $\psi_i$ is added. The choice of equilibrium distribution function $f_i^{\text{eq}}({}^0 \Pi, \vec{{}^1 \Pi}, \ten{{}^2 \Pi},...)$ and source term $\psi_i$ determines the equations modelled in moment space, as discussed below. Here and in the following, we assume a simple BKGW collision term (see \cite[p.98-101]{krueger}), and a second-order He forcing scheme \cite{he} (see \cite[p.233-239]{krueger}). 

The results of this local operation are then propagated to neighbouring lattice sites in the streaming step
\begin{equation}\label{eq:stream}
    \bar{f}_i(\vec{X}+\vec{C}_i \Delta t,t+\Delta t) = \bar{f}_i^{col}\,.
\end{equation}

The link between the distribution functions $\bar{f}_i$ and the macroscopic moments $\ten{{}^N \Pi}$, meanwhile, is given via \cite[p.239]{krueger}
\begin{equation}\label{eq:moment}
    {}^N \Pi_{[\alpha]} = \sum_i C_{i[\alpha]}^{\otimes^N} \bar{f}_i + \frac{1}{2} {}^N S_{[\alpha]} \Delta t\,.
\end{equation}
Here, the dyadic product of the velocity vector $C_i$ with itself, taken $N$ times, is denoted via a dyadic power operator $\otimes^N$ in the superscript\footnote{This dyadic power operator is defined such that $\vec{C}_i^{\otimes^1}=\vec{C}_i$, $\vec{C}_i^{\otimes^2}=\vec{C}_i \otimes \vec{C}_i$, $\vec{C}_i^{\otimes^2}=\vec{C}_i \otimes \vec{C}_i \otimes \vec{C}_i$, etc..}. For example, the first three moments are computed via
\begin{align}
    & {}^0 \Pi = \sum_i \bar{f}_i + \frac{1}{2} {}^0 S \Delta t \nonumber\,,\\
    & {}^1 \Pi_{\alpha} = \sum_i C_{i\alpha} \bar{f}_i + \frac{1}{2} {}^1 S_{\alpha} \Delta t \nonumber \,, \\
    & {}^2 \Pi_{\alpha \beta} = \sum_i C_{i\alpha} C_{i\beta} \bar{f}_i + \frac{1}{2} {}^2 S_{\alpha\beta} \Delta t\,.\nonumber
\end{align}

The moment computation by (\ref{eq:moment}) is performed after the collision and streaming steps in each iteration. The moments are then used to compute the equilibrium distribution function $f_i^{eq}({}^0 \Pi, \vec{{}^1 \Pi}, \ten{{}^2 \Pi},...)$ for use in the collision step (\ref{eq:col}) in the next iteration \cite[p.67]{krueger}. Similarly, the source term $\psi_i$ in (\ref{eq:col}) is computed as $\psi_i({}^0 S, \vec{{}^1 S}, \ten{{}^2 S},...)$ from the source terms $\ten{{}^N S}$ found on the right hand sides of the moment chain in (\ref{eq:chain}) \cite{farag}. Which conservation laws can be modelled depends on the particular choice of equilibrium distribution function $f_i^{eq}$ and source term $\psi_i$ \cite{farag}. Generally speaking, $f_i^{eq}({}^0 \Pi, \vec{{}^1 \Pi}, \ten{{}^2 \Pi},...)$ and $\psi_i({}^0 S, \vec{{}^1 S}, \ten{{}^2 S},...)$ are functions in the moments $\ten{{}^N \Pi}$ and source terms $\ten{{}^N S}$, respectively.

Finally, boundary rules must be defined to determine the value of the distribution functions $\bar{f}_i$ in accordance with the macroscopic boundary conditions on the moments $\ten{{}^N \Pi}$. In a recent preprint \cite{us}, we adapted the popular bounce-back (see \cite[p.175-181]{krueger}) and anti-bounce-back boundary rules (see \cite[p.200]{krueger}) to model the behaviour of Dirichlet and Neumann boundaries with an LBM for linear elasticity by \cite{murthy} and \cite{escande}. In this work, we use the same methodology for Dirichlet boundaries, and adjust the modified anti-bounce-back boundary rule to handle boundary conditions in the first Piola-Kirchhoff stress in the reference configuration (see section \ref{s:BC}).

\section{A moment chain for nonlinear solid mechanics}\label{s:mc}

To facilitate their solution via the LBM, we would like to rewrite the fundamental equations of solid mechanics -- the conservation of mass, the balance of linear momentum, and the material law -- in the moment chain form given in (\ref{eq:chain}). 

From the Lagrangian point of view, mass conservation becomes trivial: the reference density $\rho_0$ remains constant, and can be treated as a material parameter, i.e. $d_t \rho_0 = 0$ \cite[p.135]{holzapfel}. Thus, we do not need to account for mass conservation with the zeroth-order equation in (\ref{eq:chain}), and remain flexible as to what to model with this equation and with the (scalar) zeroth moment~${}^0 \Pi$.

The balance of linear momentum
\begin{equation}\label{eq:convmomentum}
    \rho_0 a_\alpha = d_\beta P_{\alpha \beta} + \rho_0 b_\alpha\,,
\end{equation}
meanwhile, relates the acceleration $\vec{a}$ to the divergence of the first Piola-Kirchhoff stress tensor $\ten{P}$ as well as the body force per unit mass $\vec{b}$ \cite[p.143]{holzapfel}. The acceleration $\vec{a}$ is defined as the material time derivative of the velocity $\vec{v}$, meaning that the inertial term in (\ref{eq:convmomentum}) can be rewritten as
\begin{equation}
    \rho_0 a_\alpha = \rho_0 d_t v_\alpha = d_t ( \rho_0 v_\alpha ) = d_t \bar{j}_\alpha\,,
\end{equation}
with $\vec{\bar{j}}=\rho_0 \vec{v}$ denoting the linear momentum density in the Lagrangian configuration. The balance of linear momentum thus becomes
\begin{equation}\label{eq:convmomentummod}
    d_t \bar{j}_\alpha - d_\beta P_{\alpha \beta} = \rho_0 b_\alpha\,,
\end{equation}
which is more convenient for implementation in an LB algorithm: the time derivative of the linear momentum density $\vec{\bar{j}}$ appearing here fits neatly into the first-order equation in (\ref{eq:chain}), meaning that $\vec{\bar{j}}$ might be used as a first moment $\vec{{}^1 \Pi}$.

Finally, a material law is required to account for the evolution of the first Piola-Kirchhoff stress tensor $\ten{P}$. In this publication, we assume hyperelastic behaviour of the form\footnote{Note that material law formulations in terms of alternative strain measures are also possible. The deformation gradient is chosen here because it can easily be calculated from the displacement $\vec{u}$.} $\ten{P}(\ten{H})$, with the displacement gradient $\ten{H}$
\begin{equation}\label{eq:defgrad}
    H_{\alpha \beta} = d_\beta u_\alpha\,.
\end{equation}
Here, $\vec{u}$ denotes the displacement field.
Accounting for such a material law in an LB algorithm is not trivial: ideally, we would like to use the second-order equation in (\ref{eq:chain}), i.e.
\begin{equation}
    d_t {}^2 \Pi_{\alpha \beta} + d_\gamma {}^3 \Pi_{\alpha \beta \gamma}=0\,,
\end{equation}
to model the evolution of the first Piola-Kirchhoff stress tensor $\ten{P}$ as a second moment $\ten{{}^2 \Pi}$ via the divergence of a third-order tensor $\ten{Q}$ to be used as a third moment $\ten{{}^3 \Pi}$, i.e.
\begin{equation}\label{eq:divprob}
    - d_t P_{\alpha \beta} + d_\gamma Q_{\alpha \beta \gamma}=0\,.
\end{equation}
However, an analytical solution for a third order tensor $\ten{Q}$ satisfying this relation for an arbitrary hyperelastic material law $\ten{P}(\ten{H})$ can -- to the best of our knowledge -- not be identified easily\footnote{This equation might also be solved numerically. If discretised with finite difference approximations to the derivatives, (\ref{eq:divprob}) yields a very sparse linear system. In each iteration of an LB algorithm, the value of the unknown third order tensor $\ten{Q}$ from the previous iteration might be used as an initial value for an iterative solver. Due to the small time step, the change in the values of $\ten{P}$ and thus $\ten{Q}$ from one iteration to the next is limited, promising quick convergence. Note that (\ref{eq:divprob}) can be treated separately for coordinate indices $\alpha$ and $\beta$.}. 

We instead use the second-order equation in (\ref{eq:chain}) to model a part of the material law $\ten{\bar{P}}(\ten{H})$ which is linear in the displacement gradient $\ten{H}$, i.e. 
\begin{equation}\label{eq:divproblin}
    d_t \bar{P}_{\alpha \beta} + d_\gamma \bar{Q}_{\alpha \beta \gamma}=0\,.
\end{equation}
It turns out that for the linear expression
\begin{equation}\label{eq:Qlin}
    \bar{Q}_{\alpha \beta \gamma} = C_s^2 ( \bar{j}_\alpha \delta_{\beta \gamma} + \bar{j}_\beta \delta_{\alpha \gamma} + \bar{j}_\gamma \delta_{\alpha \beta} )\,,
\end{equation}
which can readily be replicated by the LBM \cite{murthy}, $\ten{\bar{P}}$ turns out to be 
\begin{equation}\label{eq:Plin}
    \bar{P}_{\alpha \beta} = - \mu ( d_\beta u_\alpha + d_\alpha u_\beta + d_\gamma u_\gamma \delta_{\alpha \beta} )\,.
\end{equation}
Here, $C_s=\sqrt{\nicefrac{\mu}{\rho_0}}$ is the shear wave speed.
In a previous publication \cite{us}, we referred to $\ten{\bar{P}}$ as the Poisson stress tensor due to its similarity to the Cauchy stress tensor for Poisson solids\footnote{For small strains and Poisson solids with $\lambda=\mu$ ($\nu=\nicefrac{1}{4}$), $\ten{\bar{P}}=-\ten{\sigma}$, hence the name.}. For the sake of convenience, the same name will be used in the following.

The (nonlinear) remainder of the material law can then be modelled using an additional artificial source term in the balance of linear momentum, i.e.
\begin{equation}
    d_t \bar{j}_\alpha + d_\beta \bar{P}_{\alpha \beta} = \rho_0 b_\alpha + d_\beta \Big ( P_{\alpha \beta}(\ten{H})+ \bar{P}_{\alpha \beta}(\ten{H}) \Big )\,,
\end{equation}
with the modified source term $\vec{{}^1 S} = \rho_0 \vec{b} + \text{div}(\ten{P}(\ten{H})+\ten{\bar{P}}(\ten{H}))$. 

As mentioned previously, we remain flexible as to what to model with the first equation in~(\ref{eq:chain}). We are committed to $\ten{{}^1\Pi}=\vec{\bar{j}}$ by the considerations above. For simplicity, we could set ${}^0 S=0$ and obtain
\begin{equation}
    d_t r + d_\alpha \bar{j}_\alpha = 0\,,
\end{equation}
i.e.,
\begin{equation}
    {}^0 \Pi=r=-\rho_0 d_\alpha u_\alpha\,.
\end{equation}

With all the above, we have identified a moment chain formulation for nonlinear solid mechanics in the reference configuration, which is given by
\begin{align}\label{eq:chainmod}
    & d_t r + d_\alpha \bar{j}_\alpha = 0\,, \nonumber \\
    & d_t \bar{j}_\alpha + d_\beta \bar{P}_{\alpha \beta} = \rho_0 b_\alpha + d_\beta \Big ( P_{\alpha \beta}(\ten{H})+ \bar{P}_{\alpha \beta}(\ten{H}) \Big )\,, \nonumber \\
    & d_t \bar{P}_{\alpha \beta} + d_\gamma \bar{Q}_{\alpha \beta \gamma}=0\,.
\end{align}
The zeroth moment $r=-\rho_0 d_\alpha u_\alpha$ is proportional to the trace of the deformation gradient $\ten{H}$ and thus encodes first-order information about tensile and compressive strains. Its evolution is determined by the linear momentum density $\vec{\bar{j}}$, which is used as a first moment. The evolution of $\vec{\bar{j}}$ is determined by the divergence of the second moment $\ten{\bar{P}}$ and by the first order source terms on the right hand side of the second equation in (\ref{eq:chainmod}). 

The second moment $\ten{\bar{P}}$ encodes a part of the material law which is linear in $\ten{H}$. The nonlinear remainder of the material law is modelled via an additional source term -- note the opposing signs of $\ten{P}$ and $\ten{\bar{P}}$, which mean that the linear terms accounted for by $\ten{\bar{P}}$ are 'subtracted from' the first Piola-Kirchhoff tensor for the source term.

As shown by \cite{farag}, the LBM mimics a second-order Crank-Nicolson scheme for moment chains like this\footnote{For the LBM to mimic a fully implicit Crank-Nicolson solver, the source terms must be evaluated -- for some iteration starting at time $t$ -- at time $t$ for the collision operation in (\ref{eq:col}) and at time $t+\Delta t$ for the moment computation via (\ref{eq:moment}).}. The relaxation and streaming operations in an LBM algorithm account for the left hand sides of (\ref{eq:chainmod}), while the source terms -- both the bulk force and the nonlinear parts of the material law -- are modelled via the forcing contribution in the collision step.

\section{An LBM for nonlinear solid mechanics}\label{s:FSMC_LBM}

In the preceding section, we introduced a moment chain for nonlinear, large-deformation solid mechanics in the reference configuration. As moments, we selected ${}^0 \Pi=r$, $\vec{{}^1 \Pi}=\vec{\bar{j}}$, $\ten{{}^2 \Pi}=\ten{\bar{P}}$, and $\ten{{}^3 \Pi}=\ten{\bar{Q}}$, with
\begin{align}
    & \bar{P}_{\alpha \beta} = - \mu ( d_\beta u_\alpha + d_\alpha u_\beta + d_\gamma u_\gamma \delta_{\alpha \beta} )\,,\nonumber \\
    & \bar{Q}_{\alpha \beta \gamma} = C_s^2 ( \bar{j}_\alpha \delta_{\beta \gamma} + \bar{j}_\beta \delta_{\alpha \gamma} + \bar{j}_\gamma \delta_{\alpha \beta} )\,,\nonumber
\end{align}
from (\ref{eq:Plin}) and (\ref{eq:Qlin}).
These moments may readily be replicated via the LBM: the equilibrium distribution function \cite{murthy}
\begin{equation}
    f_i^{eq} = w_i \Big ( r + \frac{1}{c_s^2} C_{i \alpha} \Bar{j}_\alpha + \frac{1}{2C_s^4} ( \bar{P}_{\alpha \beta} - r C_s^2 \delta_{\alpha \beta} ) ( C_{i \alpha} C_{i \beta} - C_s^2 \delta_{\alpha \beta} ) \Big )\,,
\end{equation}
yields precisely the tensors $r$, $\vec{\bar{j}}$, $\ten{\bar{P}}$, and $\ten{\bar{Q}}$ via the moments,
\begin{align}
    r&=\sum_i f_i^{eq} \,, \nonumber\\
    \bar{j}_\alpha & = \sum_i f_i^{eq} C_{i \alpha} \,, \nonumber\\
    \bar{P}_{\alpha \beta} & = \sum_i f_i^{eq} C_{i \alpha} C_{i \beta}\,,\nonumber\\
    \bar{Q}_{\alpha \beta \gamma} & = \sum_i f_i^{eq} C_{i \alpha} C_{i \beta} C_{i \gamma}\,,\label{eq:moments}
\end{align}
which guarantees that the moments of $\bar{f}_i$ behave as in (\ref{eq:moment}) \cite{farag}.
The source term \cite{murthy},
\begin{equation}\label{eq:source}
    \psi_i = w_i \frac{1}{c_s^2} C_{i \alpha} {}^1 S_\alpha\,,
\end{equation}
with 
\begin{equation}\label{eq:source1}
    {}^1 S_\alpha = \rho_0 b_\alpha + d_\beta ( P_{\alpha \beta}(\ten{H}) + \bar{P}_{\alpha \beta}(\ten{H}) )\,,
\end{equation}
accounts for the remainder of the material law.
Since the displacement $\vec{u}$ is not directly available in the LBM algorithm, it must be computed from the linear momentum density via integration. Here, a trapezoidal rule
\begin{equation}
    u_\alpha(x,t+\Delta t) = u_\alpha(x,t) + \frac{1}{2\rho_0}\Big (\bar{j}_\alpha(x,t+\Delta t)+\bar{j}_\alpha(x,t) \Big )\,,
\end{equation}
is used.
The displacement gradient $H_{\alpha \beta}=d_\beta u_\alpha$ can then be determined from the displacement and be used to compute the first Piola-Kirchhoff and Poisson stress tensors $\ten{P}$ and $\ten{\bar{P}}$ for use in the source term $\vec{{}^1 S}$. We utilise central finite difference stencils for the gradient and divergence computations.

It is important to note that the source term $\vec{{}^1 S}$ is required at two points in each iteration: firstly, to compute the forcing contribution to the collision step (\ref{eq:col}), and secondly, in the moment computation (\ref{eq:moment}). Note that, for the LB algorithm to mimic a fully implicit Crank-Nicolson solver, $\vec{{}^1 S}$ must be evaluated at time $t$ in the collision step, but at time $t+\Delta t$ in the moment computation, for some iteration starting at time $t$.
Since the displacement at time $t+\Delta t$ is unknown prior to the moment computation, we instead evaluate $\vec{{}^1 S}$ at time $t$ in~(\ref{eq:moment}). Thus, we do not mimic a fully explicit second-order Crank-Nicolson scheme for the nonlinear component of the material law (but do for the part contained in $\ten{\bar{P}}$). The approximation error resulting from this operation is of order $\mathcal{O}(\Delta t^2)$.
If we approximate $\vec{\bar{j}}$ at $t+\Delta t$ by linear extrapolation from $t$, and subsequently use the trapezoidal rule to obtain $\vec{u}$ at $t+\Delta t$, the error reduces to $\mathcal{O}(\Delta t^3)$.
We therefore obtain a partially explicit scheme for the part of the material law not contained in $\ten{\bar{P}}$, albeit with a rather small error\footnote{If this resulted in issues with the stability of the method, we could implement a predictor-corrector scheme for the computation of $\vec{u}$ at $t+\Delta t$. In the numerical experiments performed by the authors, however, this was not necessary.}.

Note that this is a 'quick and dirty' strategy: it would be preferable, from the viewpoint of efficiency and stability, to do away with this source term, but we have not been able to do so yet. Encouragingly, the LB scheme presented here seems to perform as desired in spite of this. We expect further performance gains from a 'cleaner' algorithm.

The complete LB algorithm -- used to solve the solid-mechanical moment chain in (\ref{eq:chainRK}) -- is outlined in algorithm \ref{alg:solidlbm}. A brief discussion of the boundary handling scheme follows in the next section.

\begin{algorithm}
\SetKw{Initialise}{Initialise}
\SetKw{Set}{Set}
\Set{$\Delta t$, $\Delta X$, $C_s$, $\vec{C}_i$, $w_i$, $\bar{\tau}$, $\lambda$, $\mu$, $\rho_0$}\;
\caption{An LB algorithm for nonlinear solid mechanics.}\label{alg:solidlbm}
\Initialise{Lattice, Boundary, Moments: $r$, $\vec{\bar{j}}$, $\ten{\bar{P}}$, $\ten{\bar{Q}}$}\;
\tcp{initialise distribution functions}
$\bar{f}_i \gets w_i \Big ( r + \frac{1}{c_s^2} C_{i \alpha} \bar{j}_\alpha + \frac{1}{2C_s^4} ( \bar{P}_{\alpha \beta} - r c_s^2 \delta_{\alpha \beta} ) ( C_{i \alpha} C_{i \beta} - C_s^2 \delta_{\alpha \beta} ) \Big )$\;
$t \gets 0$\;
\tcp{main simulation loop}
\While{$t \leq t_{max}$}{
\For{$X \in L$}{
${}^1 S_\alpha \gets \rho_0 b_\alpha + d_\beta ( \bar{P}_{\alpha \beta} + P_{\alpha \beta} )$\;
\For{$i \gets 0$ \KwTo $9$}{
    \tcp{calculate equilibrium distribution function}
    $f_i^{eq} \gets w_i \Big ( r + \frac{1}{C_s^2} C_{i \alpha} \bar{j}_\alpha + \frac{1}{2C_s^4} ( \bar{P}_{\alpha \beta} - r C_s^2 \delta_{\alpha \beta} ) ( C_{i \alpha} C_{i \beta} - C_s^2 \delta_{\alpha \beta} ) \Big )$\;
    \tcp{compute source}
    $\psi_i \gets w_i \frac{1}{C_s^2}C_{i\alpha} {}^1 S_\alpha$\;
    \tcp{collide}
    $\bar{f}_i^{col} \gets \bar{f}_i - \frac{\Delta t}{\bar{\tau}} \big ( \bar{f}_i - \bar{f}_i^{eq} \big ) + \Big ( 1 - \frac{\Delta t}{2 \bar{\tau}} \Big ) \psi_i \Delta t$\;
    \tcp{stream (except for $X+C_i \Delta t$ crossing boundary)}
    $\bar{f}_i(X+C_i \Delta t,t+\Delta t) \gets \bar{f}_i^{col}(X,t)$\;
}
}
\tcp{calculate missing $\bar{f}_i$ at the boundary}
\For{$X \in \partial L_u$}{
\For{$i$ crossing $\partial \Omega_u$}{
    $\bar{f}_{\Bar{i}}(X,t+\Delta t) \gets \bar{f}_i^{col}(X,t) - \frac{2}{C_s^2} w_i C_{i \alpha} \bar{j}_\alpha^{bd}$\;
}
}
\For{$X \in \partial L_t$}{
\For{$i$ crossing $\partial \Omega_t$}{
    $\bar{f}_{\Bar{i}}(X,t+\Delta t) \gets -\bar{f}_i^{col}(X,t) + 2 w_i \Big ( r^{bd} + \frac{1}{2 C_s^4}( \bar{P}_{\alpha \beta}^{bd} - r C_s^2 \delta_{\alpha \beta} ) ( C_{i \alpha}C_{i \beta} - C_s^2 \delta_{\alpha \beta} ) \Big )$\;
}
}
\tcp{calculate moments (explicit here)}
\For{$X \in L$}{
    ${}^1 S_\alpha \gets \rho_0 b_\alpha + d_\beta ( \bar{P}_{\alpha \beta} + P_{\alpha \beta} )$\;
    $r \gets \sum_i \bar{f}_i $, 
    $\bar{j}_\alpha \gets \sum_i C_{i \alpha} \bar{f}_i + \frac{\Delta t}{2} {}^1 S_\alpha$,
}
\tcp{update displacement and stress}
\For{$X \in L$}{
    $u_\alpha(X,t+\Delta t) \gets u_\alpha(X,t) + \nicefrac{1}{2\rho_0}\Big(\bar{j}_\alpha(X,t+\Delta t)+\bar{j}_\alpha(X,t)\Big)$\;
    $H_{\alpha \beta} \gets d_\beta u_\alpha$, $P_{\alpha \beta} \gets P_{\alpha \beta}(\ten{H})$, $\bar{P}_{\alpha \beta} \gets \bar{P}_{\alpha \beta}(\ten{H})$\;
}
$t \gets t + \Delta t$\;
}
\end{algorithm}

\section{Boundary conditions}\label{s:BC}

In a recent publication \cite{us}, we adapted the the popular bounce-back (see \cite[p.175-181]{krueger}) and anti-bounce-back boundary rules (see \cite[p.200]{krueger}) to approximate the behaviour of Dirichlet and Neumann boundaries with an LBM for linear elasticity by \cite{murthy} and \cite{escande}. Here, we use an analogous methodology to impose boundary values in the linear momentum density $\vec{j}$
\begin{equation}\label{eq:bbcon}
    \bar{f}_{\Bar{i}}(\vec{x},t+\Delta t) = \bar{f}_i^{\text{col}}(\vec{x},t) - \frac{2}{C_s^2} w_i C_{i \alpha} \bar{j}_\alpha^{*}\,,
\end{equation}
and on the Poisson stress tensor $\ten{\bar{P}}$
\begin{equation}\label{eq:abbcon}
    \bar{f}_{\Bar{i}}(\vec{x},t+\Delta t) = -\bar{f}_i^{\text{col}}(\vec{x},t) + 2 w_i \Big ( r^{\text{bd}} + \frac{1}{2 C_s^4}( \bar{P}_{\alpha \beta}^{*} - r^{\text{bd}} C_s^2 \delta_{\alpha \beta} ) ( C_{i \alpha}C_{i \beta} - C_s^2 \delta_{\alpha \beta} ) \Big )\,.
\end{equation}
Boundary values in the Poisson tensor may be obtained from the boundary stresses $\vec{T^*}$ via
\begin{align}
    & P_{\alpha \beta} N_\beta = T_\alpha^* \nonumber \,, \\
    & -\bar{P}_{\alpha \beta} N_\beta + ( P_{\alpha \beta} + \bar{P}_{\alpha \beta} ) N_\beta = T_\alpha^* \nonumber \,, \\
    & \bar{P}_{\alpha \beta} N_\beta = - T_\alpha^* + ( P_{\alpha \beta} + \bar{P}_{\alpha \beta} ) N_\beta \nonumber \,.
\end{align}
As in \cite{us}, we transform these equations into a coordinate system normal to the boundary, in which the quasi-boundary tractions $-\vec{T^*}+(\ten{P}+\ten{\bar{P}})\vec{N}$ may simply be substituted into the first row and column of $\ten{\bar{P}}$, while the remaining entries are extrapolated to the boundary. The Poisson and first Piola-Kirchhoff tensors appearing in the second term are also extrapolated to the boundary.

\section{Numerical experiments}\label{s:valid}

Having proposed an LB algorithm for nonlinear solid mechanics in the previous sections, we now consider example problems for validation. 
In the absence of analytical solutions for nonlinear dynamic field problems, we instead resort to comparisons with Finite Element (FE) solutions. Firstly, two simple tension and shear load cases are investigated, and a discussion of a more challenging example follows.
For the sake of simplicity, only the two-dimensional plane strain case is treated in this publication.

The LBM is used to model the response of a compressible neo-Hooke solid to these load cases. For the two-dimensional case, the strain energy density function could be written as
\begin{equation}
    W = \frac{\mu}{2} ( I_C - 2 ) + \frac{\lambda}{4} ( J^2 - 1 ) - \frac{\lambda}{2} \log J - \mu \log J\,,
\end{equation}
and Lam\'e parameters $\lambda = \mu = 1$ as well as density $\rho_0 = 1$ are assumed throughout. Here, $\ten{C}$ is the right Cauchy-Green tensor $\ten{C}=(\ten{H}+\ten{I})^T (\ten{H}+\ten{I})$ -- of which $I_C$ is the first invariant -- and $J$ the volume ratio \mbox{$J=\text{det}(\ten{H}+\ten{I})$}. For non-dimensionalisation, we use $l_{\text{ref}}=u_{\text{ref}}=l$.

\subsection{Simple tension and simple shear}

\begin{figure}[h]
    \centering
    \begin{minipage}{0.3\textwidth}
        \centering
        \includegraphics{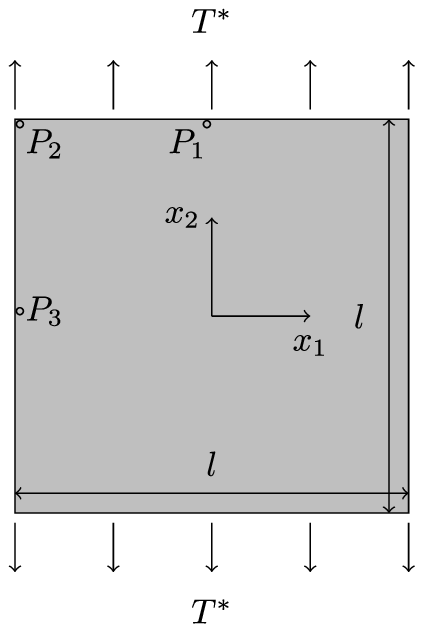}
    \end{minipage}%
    \hfill
    \begin{minipage}{0.65\textwidth}
    \centering
    \includegraphics[scale=0.9]{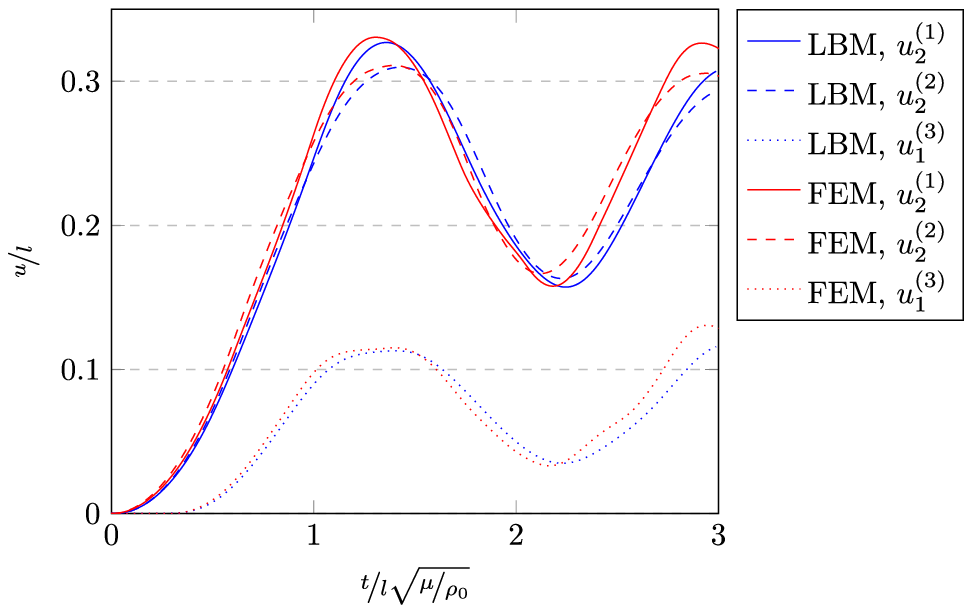}
    \end{minipage}
    \par
    \begin{minipage}{0.3\textwidth}
        \centering
        \caption{Square plate subjected to an in-plane tension load in the $x_2$-direction.}
        \label{fig:tension}
    \end{minipage}%
    \hfill
    \begin{minipage}{0.65\textwidth}
    \centering
        \caption{Displacements postprocessed from LBM (blue) and FE (red) simulations of a neo-Hooke solid under tension loading.}
        \label{fig:tension_plot}
    \end{minipage}
\end{figure}

As a simple benchmark geometry, we consider a square block with side length $l$. This block -- pictured in figure \ref{fig:tension} -- is first subjected to a pure tensile load $T^*$ in the $x_2$ direction. $T^*$ is increased linearly from $T^*=0\nicefrac{\mu l}{l}$ at $t=0 l\sqrt{\nicefrac{\rho_0}{\mu}}$ to $T^*=1\nicefrac{\mu l}{l}$ at $t=1 l\sqrt{\nicefrac{\rho_0}{\mu}}$, and then held at this level until the end of the simulation at $t=3 l\sqrt{\nicefrac{\rho_0}{\mu}}$. 

\begin{figure}[h]
    \centering
    \begin{minipage}{0.3\textwidth}
        \centering
        \includegraphics{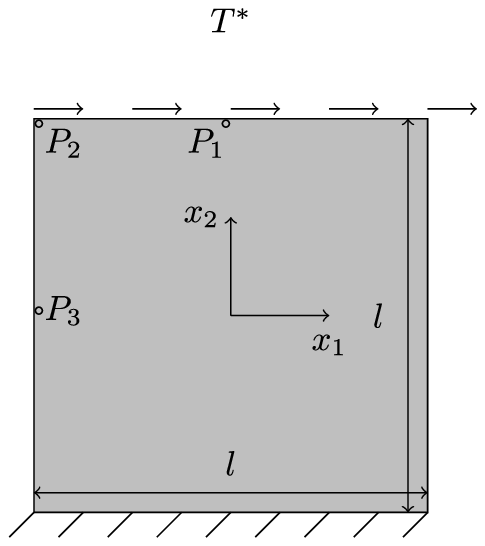}
    \end{minipage}
    \hfill
    \begin{minipage}{0.65\textwidth}
    \centering
    \includegraphics[scale=0.9]{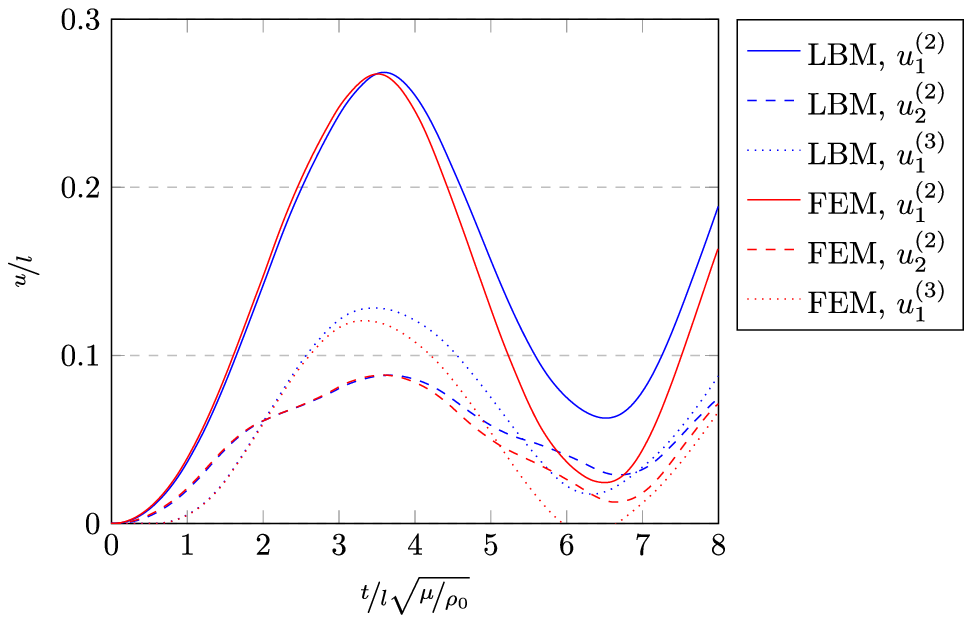}
    \end{minipage}
    \par
    \begin{minipage}{0.3\textwidth}
        \centering
        \caption{Square plate subjected to an in-plane shear load in the $x_2$-direction.}
        \label{fig:shear}
    \end{minipage}%
    \hfill
    \begin{minipage}{0.65\textwidth}
    \centering
        \caption{Displacements postprocessed from LBM (blue) and FE (red) simulations of a neo-Hooke solid under shear loading.}
        \label{fig:shear_plot}
    \end{minipage}
\end{figure}

In the second example, a pure shear load $T^*$ is applied to the same block, as displayed in figure \ref{fig:shear}. $T^*$ is increased linearly from $T^*=0\nicefrac{\mu l}{l}$ at $t=0 l\sqrt{\nicefrac{\rho_0}{\mu}}
$ to $T^*=0.05\nicefrac{\mu l}{l}$ at $t=1 l\sqrt{\nicefrac{\rho_0}{\mu}}$, and then held at this level until the end of the simulation\footnote{A reduced load is used for the shear load case due to the reduced stiffness of the block in the tangential direction. Due to the lower speed of shear waves, the simulation is run for a longer time.} at $t=8 l\sqrt{\nicefrac{\rho_0}{\mu}}$.

A lattice spacing of $\Delta X=0.025 l$ is used in both cases, yielding a time step of $\Delta t\approx 1.443\times 10^{-2} l\sqrt{\nicefrac{\rho_0}{\mu}}$ for the LBM. A FE simulation using the same material parameters is performed using FEAP \cite{feap}. 
The displacements at points $P_1=(\nicefrac{\Delta x}{2},0.5-\nicefrac{\Delta x}{2}) l$, \mbox{$P_2=(-0.5+\nicefrac{\Delta x}{2},0.5-\nicefrac{\Delta x}{2}) l$}, and $P_3=(-0.5+\nicefrac{\Delta x}{2},\nicefrac{\Delta x}{2}) l$ are then postprocessed from the simulation results produced by the LBM and the FEM.

For the simple tension simulations, figure \ref{fig:tension_plot} shows the $x_2$ component of the displacement $u_2^{(1)}$ at $P_1$, the $x_2$ component of the displacement $u_2^{(2)}$ at $P_2$, and the $x_1$ component of the displacement $u_1^{(3)}$ at $P_3$. While $u_2^{(1)}$ and $u_2^{(2)}$ are measures of the tensile response of the block in the loaded direction, $u_1^{(3)}$ quantifies the lateral contraction. As is apparent from the figure, the LBM and FEM are in agreement with regard to the computed displacements throughout the simulation, with the LBM yielding slightly lower displacements in general. Furthermore, the LBM predicts slightly later peaks in the displacement curves. While the match between LBM and FEM is not perfect, the results suggest that the LB scheme proposed above captures the large-strain tensile response of the square block in figure \ref{fig:tension} roughly as well as the FEM.

Similarly, figure \ref{fig:shear_plot} shows the $x_1$ and $x_2$ components of the displacement $u_1^{(2)}$ and $u_2^{(2)}$ at $P_2$, as well as the $x_1$ component of the displacement $u_1^{(3)}$ at $P_3$. $u_1^{(2)}$ and $u_2^{(2)}$ measure the translation of the top left corner of the block in figure \ref{fig:shear} in the horizontal and vertical directions under shear load, respectively. $u_1^{(3)}$, meanwhile, quantifies the horizontal translation of the point $P_3$ in the centre of the left side of the block. Again, the agreement between the LBM and the FEM is encouraging, though the rebound predicted by the LB scheme is less extreme. Once more, solutions computed via the LBM also peak slightly later. As both the FEM and the LBM produce only approximations to the analytical solution, it is difficult to make more precise judgements on the correctness of either result. In general, the LBM and FEM produce solutions which agree decently qualitatively and quantitatively, suggesting that the LB scheme captures a majority of the large strain shear mode response of the block in figure \ref{fig:shear}.

\subsection{Plate with hole under sudden loading and unloading}

As a slightly more challenging example, we consider a quadratic plate with side length $l$ featuring a quadratically shaped hole with side length $e=0.4l$. At $t=0 l\sqrt{\nicefrac{\rho_0}{\mu}}$, this plate is suddenly loaded with a traction of $T^*=0.1 \nicefrac{\mu l}{l}$ in the $x_2$-direction. The traction is held constant until $t=1 l\sqrt{\nicefrac{\rho_0}{\mu}}$, at which point the load is removed suddenly and $T^*=0 \nicefrac{\mu l}{l}$ until the end of the simulation at $t=3 l\sqrt{\nicefrac{\rho_0}{\mu}}$. Once more, a neo-Hookean material model is used. A lattice spacing of $\Delta X=0.0125 l$ is chosen, yielding a time step of $\Delta t\approx7.217\times 10^{-3} l\sqrt{\nicefrac{\rho_0}{\mu}}$. 

\begin{figure}[h]
    \centering
    \begin{minipage}{0.3\textwidth}
        \centering
        \includegraphics{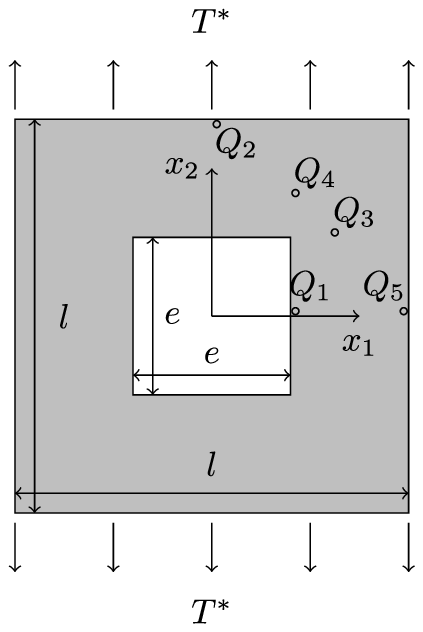}
    \end{minipage}%
    \hfill
    \begin{minipage}{0.65\textwidth}
    \centering
    \includegraphics[scale=0.9]{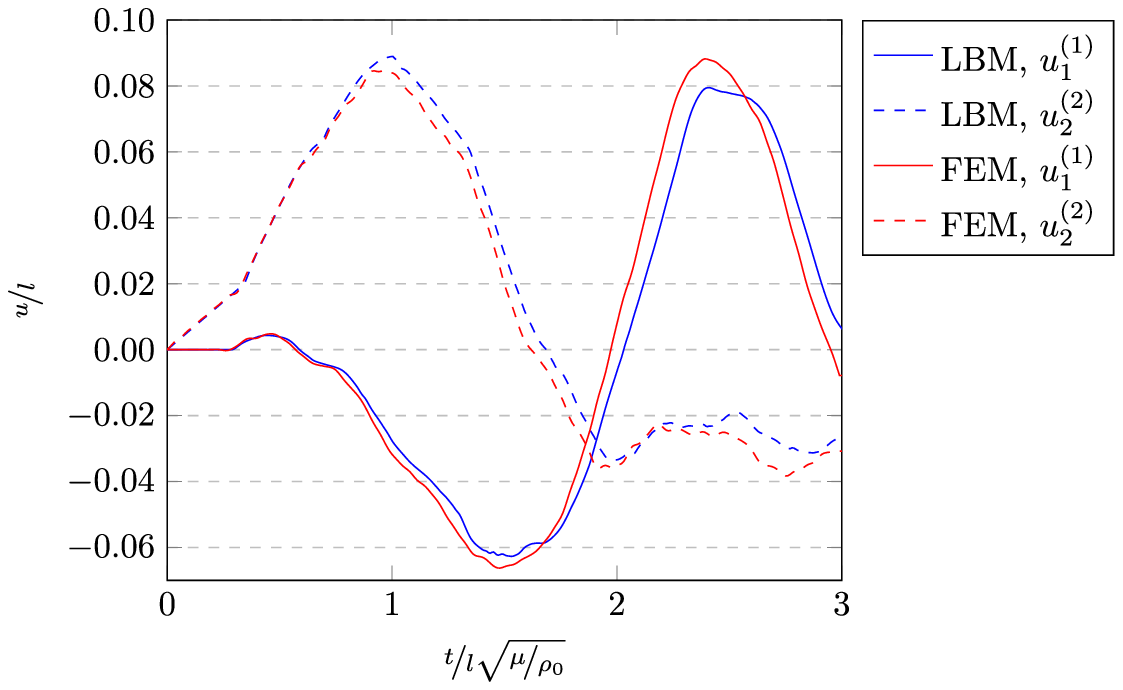}
    \end{minipage}
    \par
        \begin{minipage}{0.3\textwidth}
        \centering
        \caption{Plate with hole subjected to an in-plane tension load in the $x_2$-direction.}
        \label{fig:pwh}
    \end{minipage}%
    \hfill
    \begin{minipage}{0.65\textwidth}
    \centering
        \caption{Displacements postprocessed from LBM (blue) and FE (red) simulations of a neo-Hooke solid under tension loading.}
        \label{fig:pwh_plot}
    \end{minipage}
\end{figure}

The displacements computed by the LBM and the FEM for $Q_1=(0.2+\nicefrac{\Delta X}{2},\nicefrac{\Delta X}{2}) l$ and $Q_2=(\nicefrac{\Delta X}{2},0.5-\nicefrac{\Delta X}{2}) l$, in the $x_1$ and $x_2$ directions, respectively, are shown in figure \ref{fig:pwh_plot}. As is clear from the figure, the normal and transverse displacements computed by the FEM and LBM in response to the sudden loading and unloading are in excellent agreement throughout the simulation. The plate is deformed considerably in response to the load, being lengthened by approximately $18\%$ in the $x_2$-direction.

Furthermore, the shear component $\sigma_{12}$ of the Cauchy stress $\ten{\sigma}$ is computed at $Q_3=(0.3+\nicefrac{\Delta X}{2},0.2+\nicefrac{\Delta X}{2}) l$ and $Q_4=(0.2+\nicefrac{\Delta X}{2},0.3+\nicefrac{\Delta X}{2}) l$. Additionally, the $x_2$-direction tensile stress component $\sigma_{22}$ is evaluated at $Q_1$, $Q_3$, and $Q_5=(0.5-\nicefrac{\Delta X}{2},\nicefrac{\Delta X}{2}) l$. There is nothing fundamental about this choice of points, they are simply chosen because they yield high peak stress values and an interesting evolution of the stress in time. 

\begin{figure}
    \centering
    \includegraphics[scale=1.]{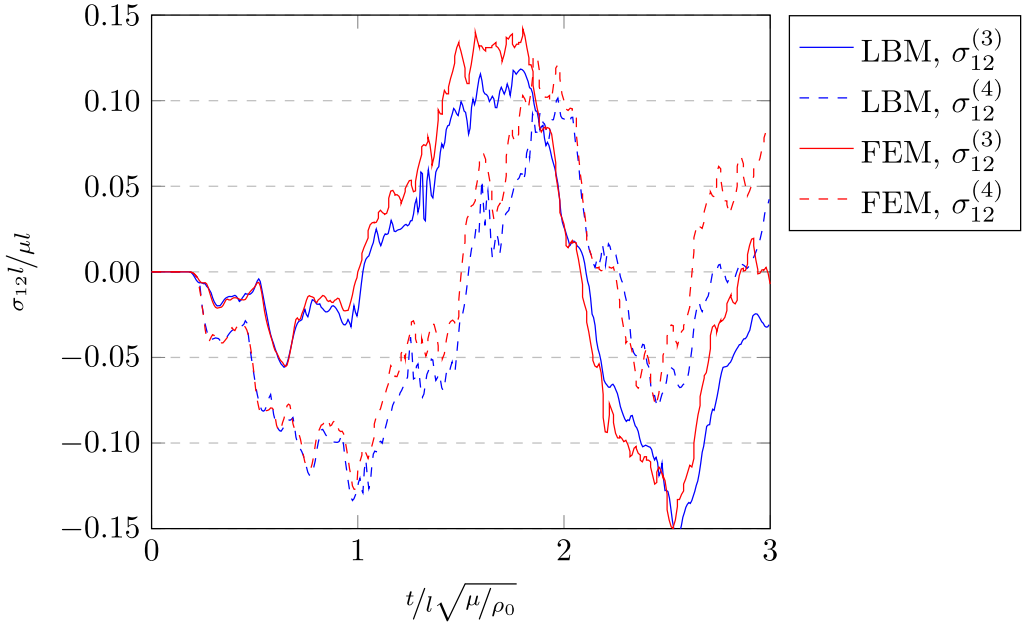}
    \caption{Shear stresses postprocessed from LBM (blue) and FE (red) simulations of a neo-Hooke solid under tension loading.}
    \label{fig:pwh_shear}
\end{figure}

The evolution of the shear stresses computed by the LBM and the FEM for the selected points is shown in figure \ref{fig:pwh_shear}. The two numerical solution match well, yielding very similar overshoot profiles. The fluctuations induced by passing S-waves are also very similar in both simulations. Generally, the LBM predicts slightly lower shear stresses throughout.

\begin{figure}
    \centering
    \includegraphics[scale=1.]{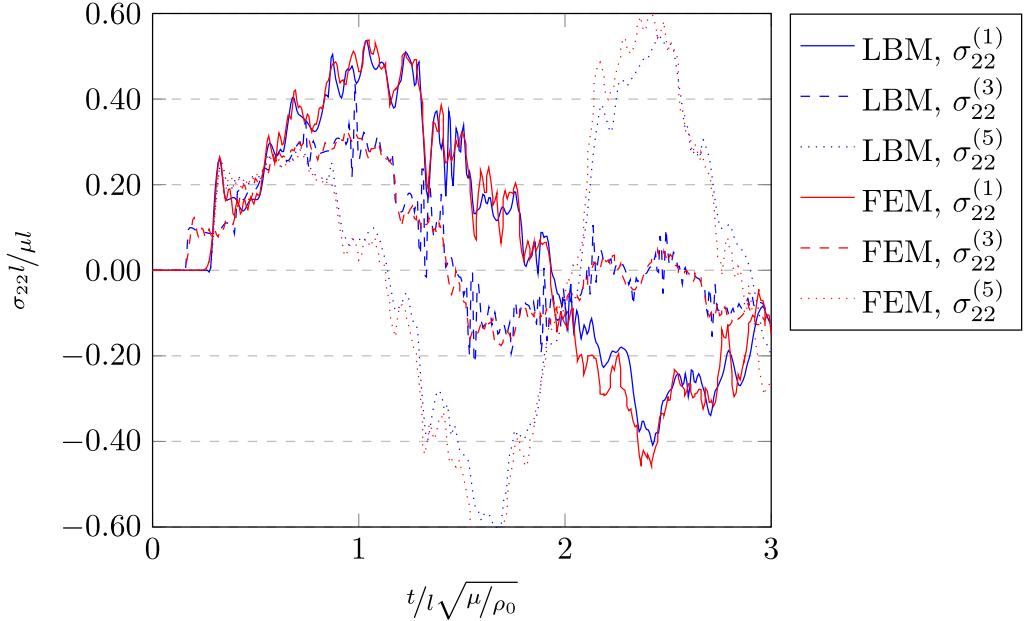}
    \caption{Tensile stresses postprocessed from LBM (blue) and FE (red) simulations of a neo-Hooke solid under tension loading.}
    \label{fig:pwh_ten}
\end{figure}

Meanwhile, figure \ref{fig:pwh_ten} outlines the evolution of the tensile stress $\sigma_{22}$, the magnitude of which is significantly higher generally. Here, the solutions obtained via the LBM and the FEM match exceptionally well. The overall overshoot profiles are nearly identical, and the sudden jumps induced by P-wave fronts in the LB and FE simulations also line up very well, with the LBM producing slightly more pronounced fluctuations overall.

\begin{figure}
    \centering
    \includegraphics[scale=0.4]{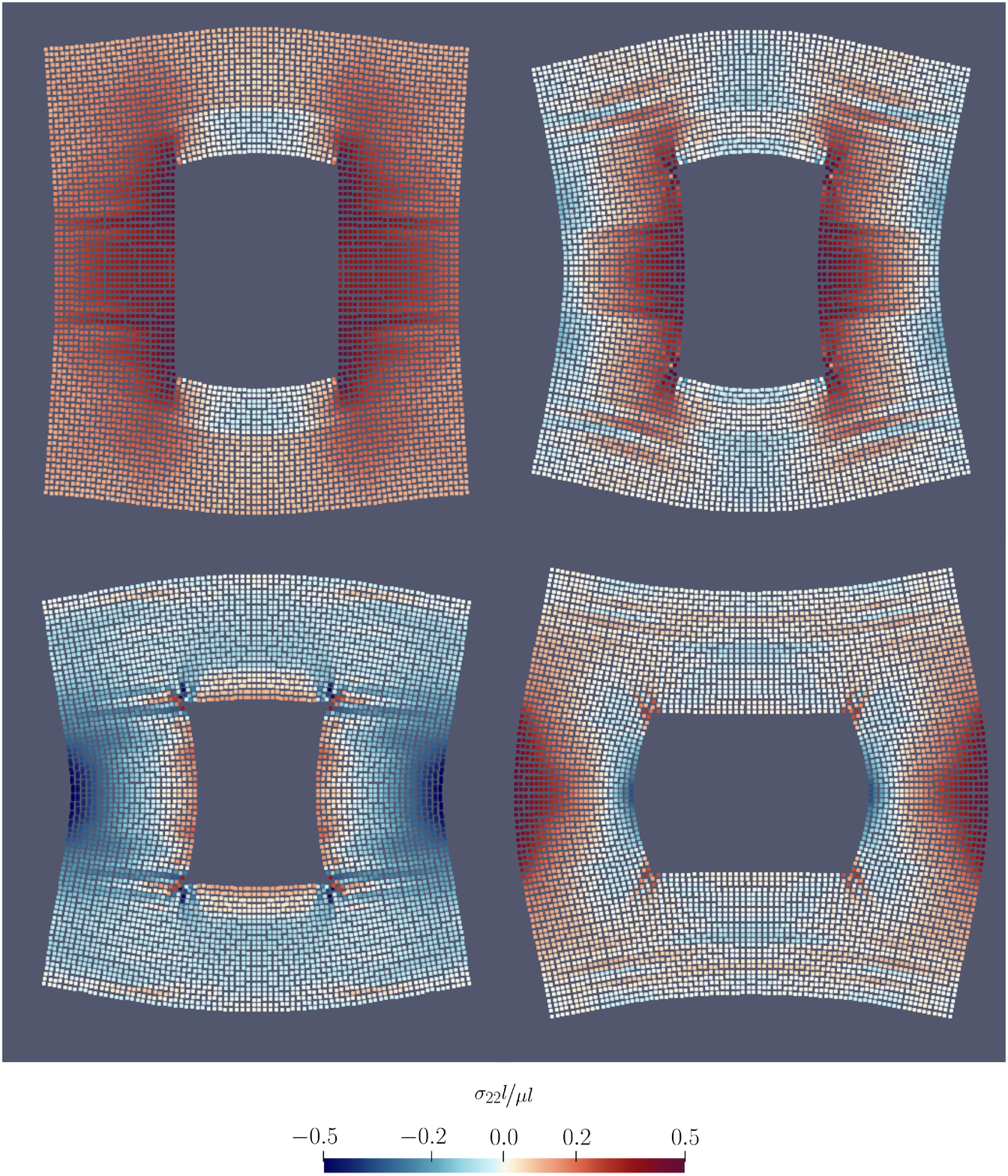}
    \caption{Deformed configurations predicted by the LBM at $t\approx0.7971 l\sqrt{\nicefrac{\rho_0}{\mu}}$ (top left), $t\approx1.232 l\sqrt{\nicefrac{\rho_0}{\mu}}$ (top right), $t\approx1.558 l\sqrt{\nicefrac{\rho_0}{\mu}}$ (bottom left), and $t\approx2.355 l\sqrt{\nicefrac{\rho_0}{\mu}}$ (bottom right). A contour plot for the $x_2$-component of the Cauchy stress $\sigma_{22}$ is superimposed on the warped domain. Deformation to scale.}
    \label{fig:defo}
\end{figure}

An animation highlighting the evolution of the deformation and the $x_2$-component of the Cauchy stress throughout the simulation is available from the authors upon request. Figure \ref{fig:defo} additionally includes four exemplary stills from this animation. The significance of the deformation -- which is shown to scale -- is apparent. Additionally, the dilatational wave fronts traversing the material domain are clearly visible. Along with the considerable dynamic overshoots in the deformation, these give rise to a complex stress field with a similarly complex time evolution.

\section{Discussion and conclusions}

Generally, FEM and LBM are in better agreement with regard to the tensile stress components than the shear stresses, both in the simple example and the more challenging benchmark. Overall, the agreement is good, with very similar peak displacement and stress values and profiles being predicted by the LBM and the FEM. 
It is difficult to asses these results in the absence of analytical benchmarks. On the other hand, the FEM is the state-of-the-art method for nonlinear, transient, large displacement solid simulation. Because of this, we think that it is encouraging that the LB algorithm presented above can replicate FE results well.
This is especially the case because the LBM for nonlinear solid mechanics is still at an early stage of development and we had to resort to artificial source terms to model part of the material law.

Additionally, the LBM seems to produce more pronounced fluctuations and jumps in the computed stresses (and, albeit to a lesser extent, in the displacements). This might be because the FEM tends to produce slight numerical damping and more stable solutions, while the LBM damps solutions to a lesser extent and is more prone to instabilities.

\section{Summary and outlook}

This work proposed a large displacement Lattice Boltzmann Method for materially and geometrically nonlinear solid mechanics in the reference configuration. 
To this end, the LBM was treated as a solver for balance laws in the reference configuration, based on a recent numerical investigation of the LBM by \cite{farag}. A moment chain for nonlinear solid mechanics was formulated, and an LB algorithm proposed to solve it.
Finally, results produced by the LBM were validated against FEM reference solutions for two simple examples and one slightly more challenging transient benchmark. 

The agreement between the LB and FE solutions is encouraging, which suggests that the LBM can indeed be used as a solver for nonlinear, transient, large displacement solid mechanics in the reference configuration. 
This is exciting, because the LB algorithm consists of computationally simple, highly parallelisable operations. An efficient implementation of the solid-mechanical LBM might have the potential to reduce simulation times for transient simulations -- for which computational efficiency is presently a bottleneck. Furthermore, the simple spatial and temporal discretisation makes for easy pre- and post-processing compared to FE workflows. 
Finally, LB schemes lend themselves to coupled simulations for fluid-structure interaction: if LB algorithms can be used on both sides of a fluid-solid interface, the required coupling might be simplified considerably.

The extension to large displacements might open up new fields of application for the LBM in sold mechanics: impact simulations and biomechanical tissue modelling, for example, may eventually become feasible. Additionally, nonlinear material models could allow for simulations of tires, seals, and polymer components under transient load. The formulation in the reference configuration, meanwhile, greatly simplifies the modelling of boundaries.

We are currently working on a three-dimensional implementation of the solid-mechanical LBM to allow for the simulation of a wider range of interesting phenomena. 
To improve the efficiency and stability of the method, we are additionally looking for a cleaner implementation of the material law which does not require source terms.
Furthermore, we would like to extend the algorithm presented in this work to inelastic material behaviour -- such as elasto-plasticity and viscoelasticity -- and to use the LBM for forming simulations.

\bibliographystyle{unsrtnat}
\bibliography{main}  






\end{document}